# Magnetic entropy change and critical exponents in double perovskite $Y_2NiMnO_6$

G. Sharma[1], T. S. Tripathi[2], J. Saha[1], and S Patnaik[1*]

[1]School of Physical Sciences, Jawaharlal Nehru University, New Delhi-110067, India.
[2]Inter-University Accelerator Centre, New Delhi-110067, India.

We report on the magnetic entropy change ($\Delta S_M$) and the critical exponents in the double perovskite manganite $Y_2NiMnO_6$ with a ferromagnetic to paramagnetic transition $T_C$ = 86 K. For a magnetic field change $\Delta H$= 80kOe, a maximum magnetic entropy change $\Delta S_M$ =-6.57J/kg-K is recorded around $T_C$. The corresponding relative cooling power (134 J/kg) is appreciable towards potential application as a magnetic refrigerant. The critical exponents, $\beta$=0.363±0.05 & $\gamma$=1.331±0.09 obtained from power law fitting to spontaneous magnetization $M_S(T)$ and the inverse initial susceptibility $\chi_0^{-1}$ satisfy well to values derived for a 3D-Heisenberg ferromagnet. The critical exponent $\delta$=4.761±0.129 is determined from the isothermal magnetization at $T_c$. The scaling exponents corresponding to second order phase transition are consistent with the exponents from Kouvel-Fisher analysis and satisfy the Widom's scaling relation $\delta$=1+($\gamma/\beta$). Additionally, they also satisfy the single scaling equation $M(H,\varepsilon) = \varepsilon^\beta f_\pm(H/\varepsilon^{(\beta+\gamma)})$ according to which the magnetization-field-temperature data around $T_C$ should collapse into two curves for temperatures below and above $T_C$.

*Corresponding Author: spatnaik@mail.jnu.ac.in

Manganite based perovskites are well-known for providing impetus to numerous technological applications that are based on magneto-electric [1,2], magneto-resistive [3], magneto-capacitive [4] and magneto-caloric effects [5]. Such juxtapositions of spin and orbital degrees of freedom are generally ascribed to double exchange [6], or inverse Dzyaloshinskii-Moriya (DM) interaction [7], but the microscopic origins are still underdebate. In particular, magnetocaloric effects, that entails change in specimen temperature due to variation in applied magnetic field has attracted considerable attention in the recent past [8]. Such magnetic refrigeration (MR) protocols are based on magnetization-demagnetization cycles and are projected to replace the existing gas compression based technology [8]. The corresponding physical parameter is the magnetic entropy change ($\Delta S_M$) which is a measure of spin disorder across a magnetic phase transition. For the success of magnetic refrigeration technology, it is imperative to find materials with large isothermal magnetic entropy change ($\Delta S_M$). This is generally achieved across first order magnetic phase transitions. Moreover, for good refrigeration, it is recommended that the material should have a large transition width and also negligible hysteretic behaviour with field and temperature cycling. This is because the relative cooling power (RCP) or cooling efficiency relates to the product of $\Delta S_M$ and full width at half maxima ($\delta T_{FWHM}$) across the magnetic transition. Notably high $\Delta S_M$ values are recently reported in materials exhibiting a first-order magnetic phase transition (FOMPT), such as $Gd_5(Ge_{1-x}Si_x)_4$ [9], $MnAs_{1-x}Sb_x$ [10], $LaFe_{13-x}Si_x$ [11], and $MnFeP_{0.45}As_{0.55}$ [12-14], etc. However, the narrow working temperature region and the presence of substantial thermal and magnetic hysteresis, often associated to the FOMPTs, limit their practical applications. In contrast, materials showing a second-order magnetic phase transition (SOMPT) usually possess a low $\Delta S_M$ but do not suffer from the aforesaid drawbacks and can be regarded as interesting candidates for the development of MR devices.

To contextualize, a direct correlation between magnetoresistance and magnetic entropy change in several perovskites [15] and chalcogenides [16] points to the fact that the double perovskites with manifold correlated ground state may be optimized as potential MR material. Specifically, the ground state magnetic order in $R_2NiMnO_6$ (R=rare earth ion) is reportedly tuneable by changing the rare earth ion [17]. Theoretical works on $Y_2NiMnO_6$ proposes it to be ferroelectric material with an intrinsic polarization in the E-type ferromagnetic ground state [17]. It also shows a giant dielectric effect at room temperature [18]. Electrical conduction has been described by polaron hopping mechanism [19]. In this communication, we identify substantial magnetocaloric characteristics in the multiferroic double perovskite $Y_2NiMnO_6$ [17]. Further, towards a microscopic understanding of such uncommon effects, we study the critical phenomena near its second order ferromagnetic transition. The corresponding critical exponents are determined from the Arrott plots [20] and are confirmed by Kouvel-Fisher method [21]. The range and dimensionality of magnetic exchange interaction are also ascertained from the scaling analysis of ln(M) Vs. ln(H) of isothermal magnetization.

Polycrystalline pellets of $Y_2NiMnO_6$ were synthesized by conventional solid state reaction method. $Y_2O_3$, NiO, and $MnO_2$ were used as starting material in stoichiometric ratio, followed by mixing and grinding. The sample was sintered in pellet form at 1180°C followed by another sintering at 1280°C post re-grinding. Room-temperature powder X-ray diffraction pattern was collected by the PANalytical X'pert PRO and diffraction pattern was analyzed by Rietveld refinement using GSAS software [22]. The magnetization measurements were performed in a *Cryogenic* physical property measurement system (PPMS).

Fig.1 shows Rietveld refinement of $Y_2NiMnO_6$ along with observed, calculated, difference and background data. Evidently, the Bragg positions are in good agreement with a monoclinic phase in the space group $P2_1/n$. Inset of Fig.1 shows schematic crystal structure of $Y_2NiMnO_6$. Magnetic ions $Ni^{2+}$ and $Mn^{4+}$ appear alternately along **c**-axis. The octahedral polyhedras of $NiO_6$ and $MnO_6$ are tilted and exhibit planar periodicity [23]. This type of tilted ordering is favoured when there is dissimilar valencyin the two transition metal ions. The $Y^{3+}$ ion is interspaced between the planar layers containing octahedral $NiO_6$ and $MnO_6$. The refined lattice parameters are estimated to be **a**=5.224Å, **b**=5.544Å, **c**= 7.481Å, α=90°, β=89.77° and $\gamma$ = 90°. The atomic parameters, bond angle and bond lengths between selected ions are listed in Table 1.

We note that the average bond length of $Ni^{2+}$-O and $Mn^{4+}$-O is different and that implies their different ionic states ($Ni^{2+}$ and $Mn^{4+}$) [23-25]. Bond angle between $Ni^{2+}$ - O - $Mn^{4+}$ are, 142.2(2)°, 146.8(1)° and 141.9(1)° that relate to the three in-equivalent oxygen atoms $O_1$, $O_2$, and $O_3$ respectively. Oxygen atoms $O_1$ occupy apical positions of polyhedra and $O_2$, and $O_3$ are in-plane. We note that the bond angles are slightly greater than a sister compound $Y_2CoMnO_6$ where the antiferro-magnetic ordering occurs at ~ 80K [24]. Bond angles ranging from 141.9° to 146.8° in $Y_2NiMnO_6$ is a signature of the cooperative tilting of adjacent $Ni^{2+}/Mn^{4+}O_6$ octahedras that results in distortion of the monoclinic unit cell.

To ascertain the magnetic state, moment per formula unit (f.u.), and magnetic transition temperature, in Fig.2(a)-(b) we have shown the change in magnetization of the sample as a function of temperature and external magnetic field. Fig.2(a) shows the isothermal magnetization Vs. magnetic field, M(H) at T = 5K with fields up to 80 kOe. The inset of Fig.2(a) shows hysteresis loop of the M(H) data on an expanded scale for low fields reflecting a soft ferromagnetic state. Fig.2(b) presents the field cooled (FC) and zero field cooled (ZFC) magnetization as a function of temperature. Further, the inverse susceptibility as a function of temperature, , $\chi^{-1}$(T) for $\mu_oH$ = 10 kOe are plotted against right y-axis. The inset in Fig.2(b) presents the M(T) as well as the temperature derivative, dM/dT for 100 Oe field of the FC data. The M(H) isotherm at 5K shows low coercive field of 150 Oe and remnant magnetization 0.52 $\mu_B$/f.u.. As expected for a soft ferromagnet, M(H) shows a rapid increase at low fields (<1000 Oe), followed by a slow variation that saturates at higher fields. The saturation magnetization ($M_{sat}$) ~ 4.52$\mu_B$/f.u. obtained from the M(H) isotherm at 5 K, is consistent with the spin only value of 5$\mu_B$/f.u. for $Ni^{2+}$:($3d^8$) and $Mn^{4+}$:($3d^3$) configuration configuration. The M(T) curves show a ferromagnetic to paramagnetic (FM-PM) transition with rapid decrease of magnetization for both ZFC and FC conditionswith increasing temperature. The sharp negative peak of dM/dT at T=85K is identified as the ferromagnetic Curie temperature ($T_C$). It is observed that at low field (100 Oe) the sample has a large bifurcation in M(T) under FC and ZFC condition that is suppressed at higher field of 10 kOe. This is a signature of spin glass like behaviour [26]. The inverse magnetic susceptibility fits well to the Curie-Wiess law $\chi(T)=C/(T-\theta_P)$ where C is the Curie constant and $\theta_P$ is the paramagnetic Curie-Wiess temperature. A deviation from the linear Curie-Weiss fit for $\chi(T)$ at 10 kOe, is observed for temperatures below 120 K. This suggests the presence of short-range weak spin correlations just above the $T_C$. These spin correlations above $T_C$ are responsible for the large bifurcation in M(T) in the presence of low external field. The paramagnetic parameters C and $\theta_P$ for 10 kOe magnetic fields are 3.40 emu-K/Oe-mole and 107.38 K respectively. The effective magnetic moments in the paramagnetic state at 10 kOe is 5.21 $\mu_B$/f.u. which is in agreement with theoretical magnitude 5.9 $\mu_B$.

Towards estimating the magnetic entropy change across the ferromagnetic transition, in Fig.3(a) we show the isothermal magnetization Vs. magnetic field (upto 80 kOe) at different temperatures from 71 K to 115 K with a temperature interval $\Delta T=2K$. Below $T_C$ the curves show a rapid increase of M for low fields followed by slow increase or saturation for higher fields. Above $T_C$, magnetization increases linearly with magnetic field. These curves have been used for analysing the critical behaviour as well as to calculate the magnetic entropy change of the sample. The magnetic entropy change is estimated from the M(H) isotherms in Fig.3(a), using the Maxwell's relation:

$$\Delta S_M = S(T,H) - S(T,0) = \mu_0 \int_0^H \left(\frac{\partial M}{\partial T}\right)_H dH \qquad 1.$$

In Fig.3(b) we plot the magnetic entropy change -$\Delta S_M(T)$, under different constant magnetic fields. We note that -$\Delta S_M(T)$ shows a broad peak centred at $T_C$ that decreases symmetrically on either side as expected for a second order FM-PM transition. The peak height increases with increasing H and the peak position does not shift in the temperature scale with magnetic field. The maximum value -6.57 J/kg-K is obtained for $\Delta H$ = 80 kOe. This value of -$\Delta S_M$ is large and is comparable to polycrystalline manganite (4.33 J/kg K) [15] and chalcogenides (1.7 J/kg K) [16]. Further with $\delta T_{FWHM}$ = 40K, cooling efficiency RCP is estimated to be 134 J/kg. The large magnetic entropy change and absence of hysteresis in M(T) at high fields make this compound important for potential applications involving magneto-caloric effects particularly around liquid nitrogen temperatures.

To get a deeper understanding about the magnetic phase transition that relates to large change in magnetic entropy, in Fig.4 we show modified Arrott plots for different values of critical exponents β and γ. The critical exponents and critical temperature are determined by analysing the Arrott plot at temperatures around $T_C$ [21,27,28]. From the Landau theory of phase transitions, the Gibbs free energy G for FM-PM transition can be expressed as

$G(T,M) = G_0 + aM^2 + bM^4 - MH.$ (2)

where the equilibrium magnetization, M, is the order parameter and the coefficients a and b are temperature-dependent coefficients [29]. At equilibrium ∂G/∂M=0, (i.e., energy minimization), and the magnetic equation of state can be expressed as

$H/M = 2a + 4bM^2.$ (3)

Thus, the plots of $M^2$ Vs. H/M should appear as parallel straight lines for different temperatures above and below $T_C$ in the high-field region. The intercepts of $M^2$ on the H/M axis is negative or positive depending on phenomena below or above $T_C$ and the line at T= $T_C$ passes through the origin. According to the criterion proposed by Banerjee [30], the order of the magnetic phase transition can be determined from the slope of the straight line; the positive slope corresponds to the second-order transition while the negative slope relates to the first-order transition. The positive slope of straight lines of $M^{1/\beta}$ Vs. $(H/M)^{1/\gamma}$ plot in Fig.4(a) for $\beta = 0.5$ and $\gamma = 1.0$ indicates that the PM-FM phase transition in this compound is a second order phase transition. However, the curves are not linear and show an upward curvature even in the high-field region, which indicates that the mean-field mean- field $\beta = 0.5$ and $\gamma = 1.0$ are not satisfied according to the Arrott-Noakes equation of state $(H/M)^{1/\gamma} = (T - T_C)/T_C + (M/M_1)^{1/\beta}$ [27]. Here M1 is a parameter to make the factor dimensionless. In other words, the Landau theory of phase transition or the mean-field theory with $\beta = 0.5$ and $\gamma = 1.0$ is not valid for this compound. Thus, to obtain the correct values of $\beta$ and $\gamma$ the modified Arrott plots, figure 4(b)-(d), are employed.

Further, according to the scaling theory, the critical exponents, $\beta$ and $\gamma$, can be estimated by approaching $T_C$ from either side of the transition while the critical exponent $\delta$ can be obtained from the critical isotherm at $T = T_C$ [27,28] The exponent $\delta$ is related to both the exponent $\beta$ and $\gamma$ via Widom's relation $\delta = 1 + (\gamma/\beta)$. Mathematically, the critical exponent from magnetization can be described as [27,28]:

$$M_S(T) = M_0|\varepsilon|^\beta, \quad \varepsilon < 0 \text{ for } T < T_C \qquad 4.$$

$$M = DH^{1/\delta}, \quad \varepsilon = 0 \text{ for } T = T_C \qquad 5.$$

$$\chi_0^{-1} = (h_0/M_0)\varepsilon^\gamma, \quad \varepsilon > 0 \text{ for } T > T_C \qquad 6.$$

Where $\varepsilon = (T - T_C)/T_C$ is the reduced temperature and $M_0$, $h_0$ and D are critical amplitudes. The exponent $\beta$ is associated with the spontaneous magnetization ($M_S$)) (i.e. in absence of magnetic field) below $T_C$ while the exponent $\gamma$ relates to the initial inverse magnetic susceptibility ($\chi_0^{-1}$) above $T_C$.

In theory, there are three kinds of trial exponents that are used to plot the modified

Arrott plots. Namely, the 3D-Heisenberg model (β =0.365, γ =1.336), the 3D-Ising model β =0.325, γ =1.24) and the tri-critical mean field model β =0.25, γ =1.0) that are employed in Fig.4(b)-(d). In all the three models, the lines in the high magnetic field region are parallel but there are deviation from linearity at the low $(H/M)^{1/\gamma}$ values. Clearly, the tri-critical mean-field model can be safely excluded because these curves show diverging trends at high fields. The 3D-Ising and the 3D Heisenberg model in Fig.4(b) and Fig.4(c) are quite similar. Thus, to confirm the better model to fit our experimental data, we have defined the relative slope (RS) =Slope at T/Slope at $T_C$(= 85 K) which defines the degree of parallelism among the lines at different temperatures with respect to the line at $T_C$. The RS Vs. T plots for all the models are shown in the inset of Fig.4(d). As is evident from the definition of RS, the most appropriate model should be the one having RS close to one for all temperatures. This is because the modified Arrott plots are a series of parallel lines [27,28]. From the above analysis, we conclude that the 3D-Heisenberg model is most appropriate to explain exchange phenomena in $Y_2NiMnO_6$. This is in contrast to the conclusions of a recent theoretical paper by Sanjeev Kumar et al. [17] on the same compound where it was treated as a 1D Ising model. To obtain the more accurate values of β and γ we have plotted the temperature dependence of the spontaneous magnetization $M_S(T, 0)$ and the inverse initial susceptibility $\chi_0^{-1}$ (T, 0) as a function of temperature in Fig.5(a). The best fits to the experimental data according to Eq. (4) and Eq. (6) yield the new critical exponents β =0.363±0.036 with $T_C$ = 86.70±0.23 and andγ =1.331±0.15 with $T_C$ = 85.18±0.35, respectively. The spontaneous magnetization $M_S(T, 0)$ and the inverse initial susceptibility $\chi_0^{-1}$ (T, 0) data can be obtained from the intercepts with the $(H/M)^{1/\gamma}$ and $M^{1/\beta}$ axes for 3D-Heisenberg model by linear extrapolation of the lines from the high-field region. The intercept on the $(H/M)^{1/\gamma}$ axis yields the $\chi_0^{-1}$(T, 0) while the intercept on $M^{1/\beta}$,$M_S(T, 0)$. The third critical exponent δ can be determined using Eq. (5). The ln(M)Vs.ln(H) plot in the high field region should be a straight line with slope $1/\delta$. The isothermal magnetization at $T_C = 85$ K is given in Fig.5(b), and the inset of Fig.5(b) shows the same on a log-log scale. The ln(M) vs ln(H) plot yields the third exponent $\delta = 4.31 \pm 0.1$ for the magnetic fields H >10 kOe. This value of $\delta$ matches well with the value of $\delta = 4.66$ as obtained from the Widom's scaling relation $\delta = 1+\gamma/\beta$ [21,27,28]. Overall, this proves that the obtained critical exponents are consistent.

Alternatively, the critical exponents can also be determined from the Kouvel-Fisher (KF) method [21]:

$$\frac{M_S(T)}{dM_S(T)/dT} = \frac{(T-T_C)}{\beta} \qquad 7.$$

$$\frac{\chi_0^{-1}}{d\chi_0^{-1}/dT} = \frac{(T-T_C)}{\gamma} \qquad 8$$

According to Eq.(7) and Eq.(8), the plot of the plot of $\frac{M_S(T)}{dM_S(T)/dT}$ Vs. T and $\frac{\chi_0^{-1}}{d\chi_0^{-1}/dT}$ Vs. T should be straight lines with their slopes $1/\beta$, $1/\gamma$ and intercepts $-T_C/\beta$ and $-T_C/\gamma$ respectively. As shown in Fig.5(c) the best fit to these straight lines yield the exponent $\beta = 0.375 \pm 0.033$ with $T_C = 85.53 \pm 0.34$ and $\gamma = 1.331 \pm 0.09$ with $T_C = 84.59 \pm 0.28$. In agreement with the analysis based on Landau theory, the KF critical exponents are consistent with that obtained from the modified Arrott plot of the 3D-Heisenberg model.

For further confirmation, the critical exponents are tested against the predictions of the scaling hypothesis. According to scaling hypothesis, in the critical region, the magnetization can be written as [27,28].

$$M(H,\varepsilon) = \varepsilon^\beta f_\pm(H/\varepsilon^{(\beta+\gamma)}) \qquad (9)$$

Where $f_\pm$ are regular functions with $f_+$ for T >$T_C$, and $f_-$ for T <$T_C$. According to predictions of the scaling hypothesis the M(H, $\varepsilon$)$\varepsilon^{-\beta}$ Vs. H$\varepsilon^{-(\beta+\gamma)}$ plots should form two universal curves for T >$T_C$ and T <$T_C$, respectively. In Fig.5(d) we have plotted the isothermal M(H, $\varepsilon$)$\varepsilon^{-\beta}$ Vs. H$\varepsilon^{-(\beta+\gamma)}$ curves for four temperatures around $T_C$ with $\beta$ and $\gamma$ values as obtained from KF analysis. The inset of the same figure shows the data on log-log scale. As expected from scaling hypothesis, all experimental data diverge into two separate set of curves appropriate for temperatures below and above $T_C$.

In conclusion, we report magneto-caloric nature and detailed critical exponent analysis via thermodynamic probing of a novel double perovskite $Y_2NiMnO_6$. This class of compounds are recently confirmed to exhibit multiferroicity driven by magnetostriction. Our study suggests that $Y_2NiMnO_6$ undergoes a second order paramagnetic to ferromagnetic transition at 85K under the gamut of 3-D Heisenberg model with the best value critical exponents $\beta=0.365$, $\gamma=1.336$ and $\delta=4.76$. Using a variety of analytical methods such as Kouvel-Fisher, Widom's and Mean-Field scaling hypothesis, we prove self-consistency in the critical exponents. A large change in magnetic entropy due to adiabatic magnetization is observed that augers well for its potential application as magnetic refrigerant around liquid nitrogen temperature.

Advanced Instrument Research Facility (AIRF), JNU is acknowledged for access to PPMS. We acknowledges A. K. Rastogi for valuable discussion. GS, TST, and JS acknowledge University Grant Commission (UGC) and Council for Scientifc and Industrial Research (CSIR), India respectively for financial support.

**Table 1.** Room temperature atomic fractional coordinate, bond length and bond angle of $Y_2NiMnO_6$.

| Atomic Positions | | | | Bond Length | | Bond Angle | |
|---|---|---|---|---|---|---|---|
| Atom | X | Y | Z | Bond | Distance(Å) | Bond | Angle(degree) |
| Y | 0.5180(18) | 0.5683(10) | 0.2497(30) | Ni-O1 | 2.00(5) | Ni-O1-Mn | 142.2(2) |
| Ni | 0.5000 | 0.0000 | 0.0000 | Ni-O2 | 2.00(6) | | |
| Mn | 0.0000 | 0.5000 | 0.0000 | Ni-O3 | 1.977(0) | Ni-O2-Mn | 146.8(1) |
| O1 | 0.388(8) | 0.952(8) | 0.253(8) | Mn-O1 | 1.96(6) | | |
| O2 | 0.197(1) | 0.206(12) | -0.058(8) | Mn-O2 | 1.97(6) | Ni-O3-Mn | 141.9(1) |
| O3 | 0.3228 | 0.6953 | -0.0593 | Mn-O3 | 2.051(6) | | |

**Figure Captions**:

FIG. 1. Room temperature of X-ray diffraction pattern of $Y_2NiMnO_6$. Solid circle, red line, green line, blue line, and magenta bar mark represent observed, calculated, background, difference between observed and calculated and Bragg position respectively. Inset is showing monoclinic crystal structure in which octahedral coordination of $Ni^{2+}$ (green polyhedra), and $Mn^{4+}$ (blue polyhedra) appear alternately along c-axis.

FIG 2. (a) Isothermal magnetization versus magnetic field, M(H) at 5 K. Inset shows the hysteresis loop for small fields on an expanded scale. (b) The zero field-cooled and field-cooled magnetization versus temperature curves for two fields, 100 Oe and 10 kOe. Inverse susceptibility versus temperature, $\chi^{-1}(T)$ is scaled in right y-axis. The inset shows the M(T) data for 100 Oe field on left y-axis and the temperature derivative of magnetization, dM/dT on right y-axis.

FIG 3. (a) Isothermal magnetization versus magnetic field from 71 K to 115 K with a temperature interval $\Delta T = 2$ K are shown. (b) Magnetic entropy change $-\Delta S_M(T)$ is plotted as function of temperature around transition temperature $T_C$ at fixed magnetic fields ranging from 0.5T - 8.0T.

FIG 4. Isothermal $M^{1/\beta}$ versus $(H/M)^{1/\gamma}$ plots at several temperatures around $T_C$ for $Y_2NiMnO_6$ for different values of β and γ. (a) Arrott plot (β =0.5, γ =1.0), (b) 3D-Heisenberg model (β =0.365, γ =1.336), (c) 3D-Ising model (β =0.325, γ =1.24) and (d)Tri-critical mean-field model (β =0.25, γ =1.0). The inset of (d)shows the relative slope(RS) as a function of temperature defined as the ration of slope at T to slope at $T_C(= 85$ K).

FIG 5. (a)Temperature dependence of the spontaneous magnetization $M_S(T)$ (left y-axis) and the inverse initial susceptibility $\chi_0^{-1}(T)$ (right y-axis). The solid curves show the best fit based on power law.(b)The isothermal M(H) at T = 85 K close to $T_C$ to calculate the critical exponent δ. The inset shows the same plot of log-log scale and the solid line is the linear fit following Eq. (4) for fields greater than 10 kOe. (c) Kouvel-Fisher plot for the spontaneous magnetization $M_S(T)$ (left y-axis) and the inverse initial susceptibility $\chi_0^{-1}(T)$ (right y-axis). Solid lines are the linear fit to the same to determine the critical exponents β & γ. (d) Scaling plot below and above $T_C$ using β and γ determined from the Kouvel-Fisher method. The inset shows the same on the log-log scale.

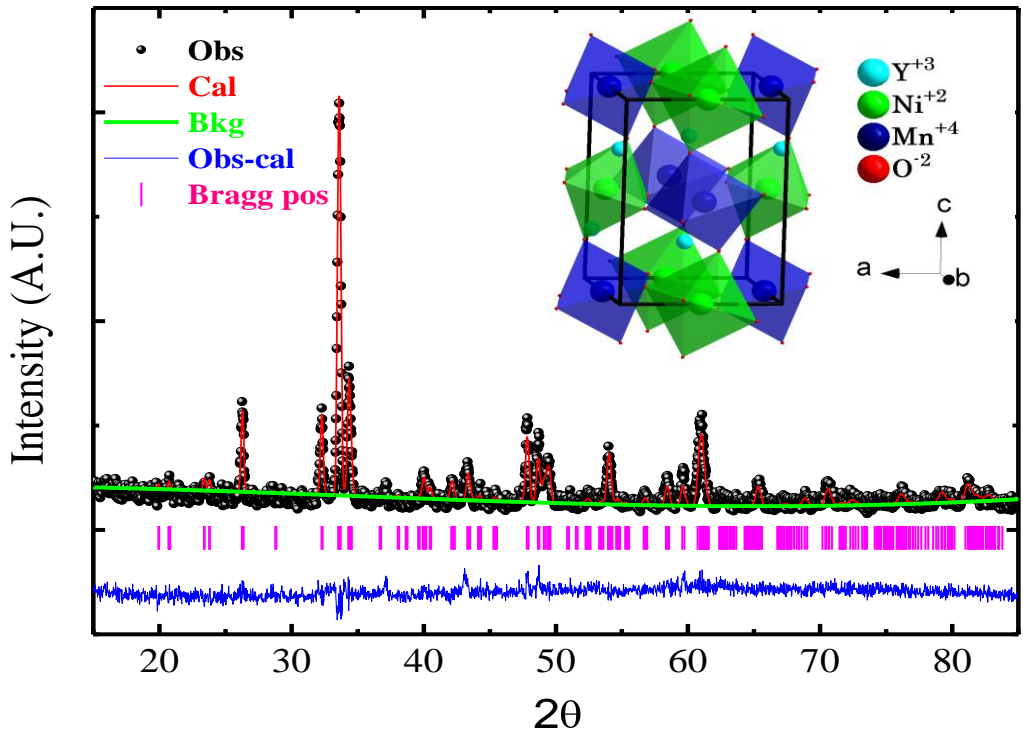

**Figure 1**

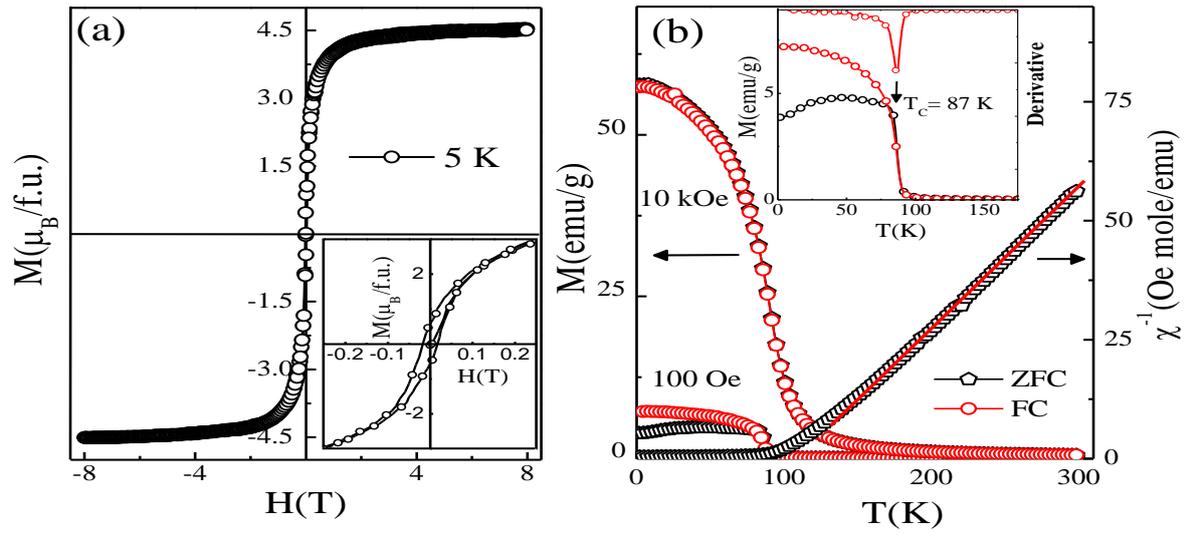

**Figure 2**

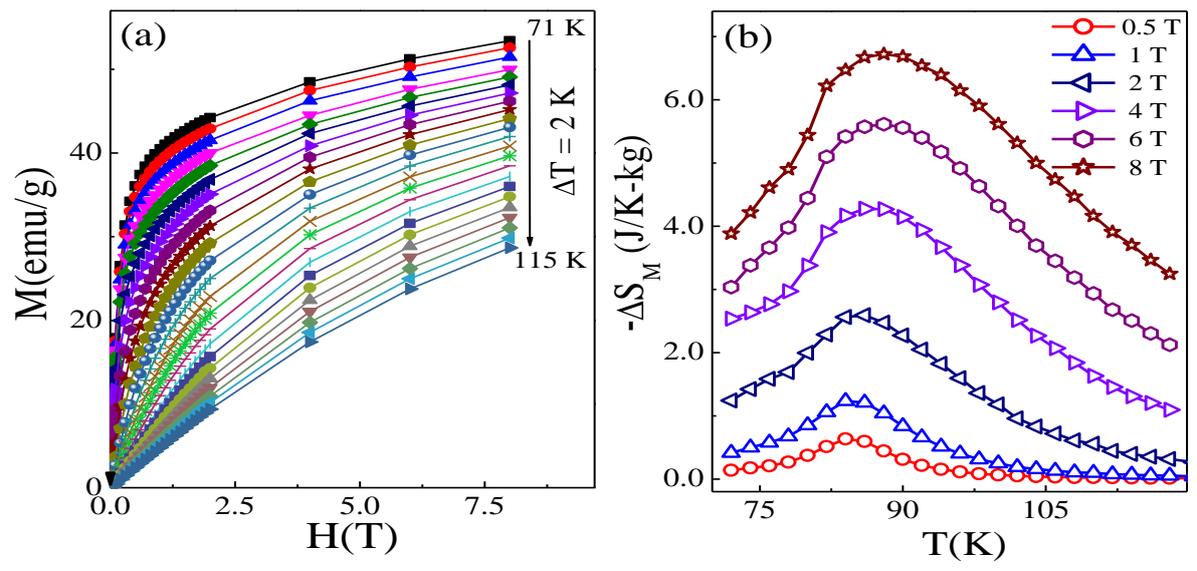

**Figure 3**

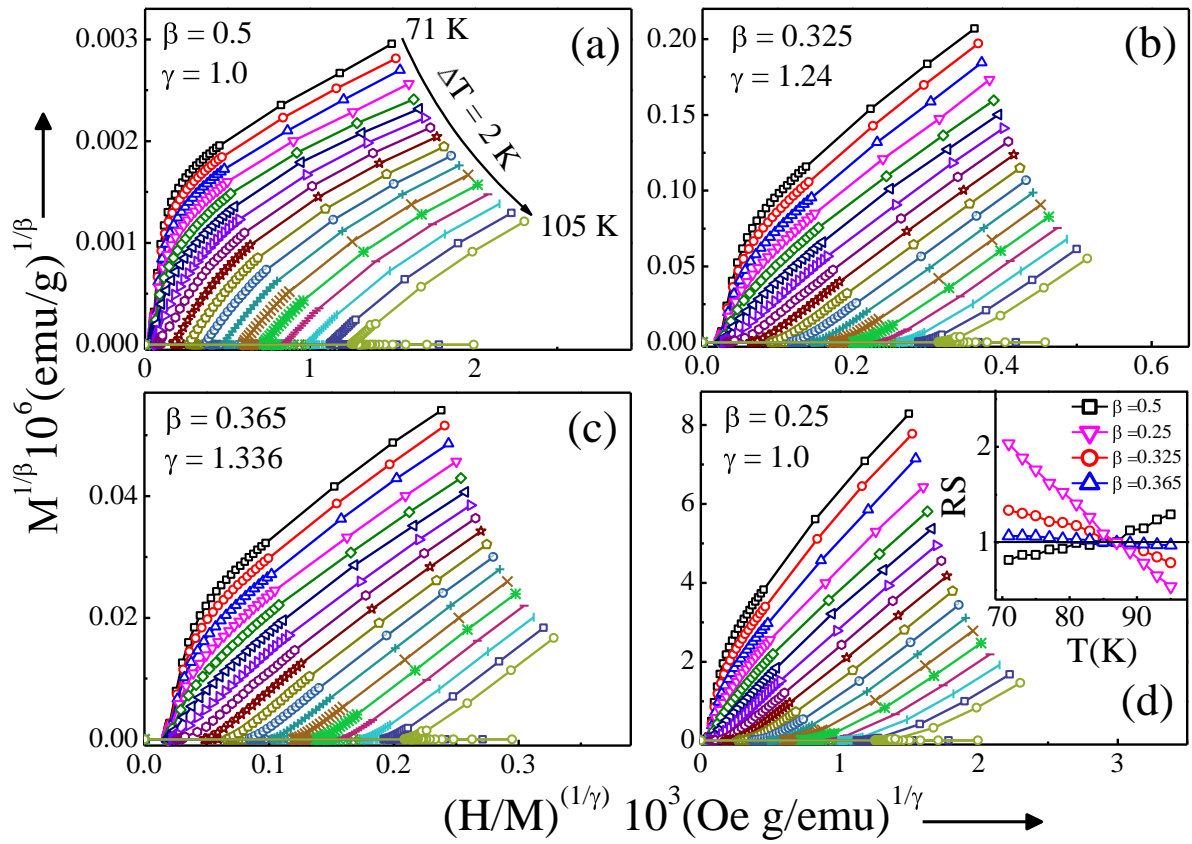

Figure 4

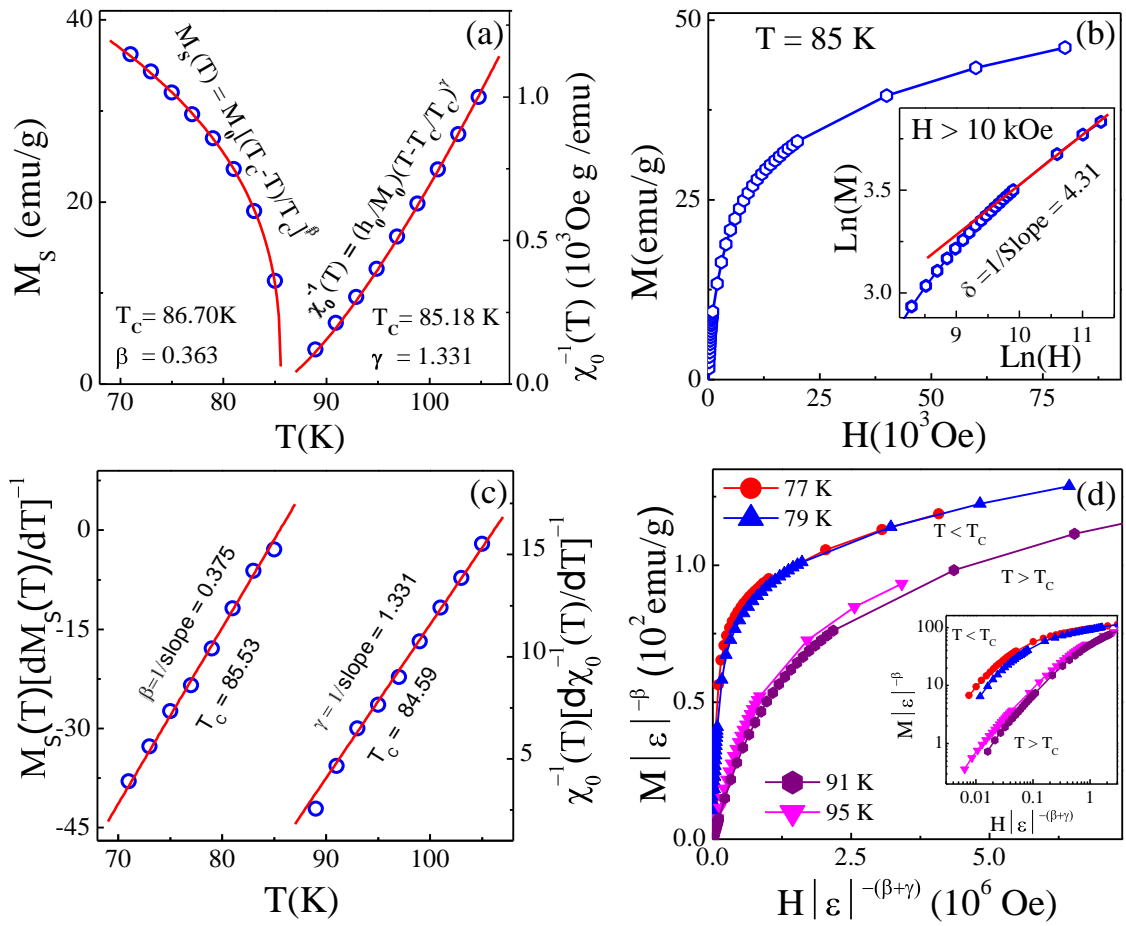

Figure 5